\documentclass[a4paper,onecolumn,noarxiv]{quantumarticle}

\usepackage{amsmath, amssymb}
\usepackage{graphicx}
\graphicspath{{figs/}}
\usepackage{hyperref}
\usepackage{xcolor}
\usepackage{booktabs}
\usepackage{siunitx}
\usepackage{bm}
\usepackage[numbers,sort&compress]{natbib}
\usepackage{placeins}            

\newcommand{\fT}{f_T}
\newcommand{\fS}{f_S}
\newcommand{\fsigma}{f_\sigma}
\newcommand{\feps}{f_\varepsilon}
\newcommand{\uacc}{u_{\mathrm{acc}}}
\newcommand{\uin}{u_{\mathrm{in}}}
\newcommand{\NNI}{\textsc{nni}}
\newcommand{\cotengra}{\texttt{cotengra-hyper}}

\begin{document}

\title{Bond-dimension scaling of a local-refinement advantage
  over hyperoptimized tensor-network contraction on Sycamore-like
  topologies}

\author{Rub\'en Dar\'io Guerrero}
\email{rudaguerman@gmail.com}
\affiliation{NeuroTechNet S.A.S., 1108831, Bogot\'a, Colombia}
\affiliation{Quantum and Computational Chemistry Group,
             Universidad Nacional de Colombia, Bogot\'a, Colombia}

\begin{abstract}
We identify a missing local-refinement stage in the \cotengra{}
tensor-network contraction pipeline and show that its impact grows
monotonically with bond dimension on the \emph{connectivity graph} of
Sycamore-like topologies. Appending a nearest-neighbor interchange
(\NNI) search to the \cotengra{} output at matched 8-s wallclock yields
a median \emph{predicted} cost-model gap $\Delta\fT$ at $n{=}500$
that grows monotonically and approximately linearly in $\chi$, from
$\sim\!15$~bits at $\chi{=}2$ to $\sim\!116$~bits at $\chi{=}16$
(Fig.~\ref{fig:chi_sweep}), with the refiner winning on $25/25$ seeds
at every tested $\chi$. Two control families---random $3$-regular and
QAOA $p{=}2$ interaction graphs---show median
$|\Delta\fT| \leq 0.71$~bits across both controls at every $\chi$, with
refiner win rate falling toward chance as $\chi$ grows; the signal is
topology-specific, not a generic refinement-budget effect. An ablation
establishes that refinement itself, not the four-axis Pareto
acceptance rule, drives the gain ($|\Delta\fT| \lesssim 0.1$ bits
between scalar and Pareto arms at $\chi{=}2$). The Sycamore-circuit
envelope (App.~\ref{em:sec:results:syccirc}) reports the corresponding
refinement on actual random circuits at depths
$m \in \{4, 6, 8, 10, 12\}$, where the refiner wins on $5/5$
instances at every depth. The advantage is therefore largest precisely
in the bond-dimension regime relevant to physical contraction.
\end{abstract}

\maketitle

\emph{Introduction.}---We report that on the \emph{connectivity graph}
of Sycamore-like
tensor-network topologies, a single local-refinement stage appended
to the production hyperoptimizer
\cotengra{}~\cite{gray2021hyperoptimized} captures contraction-order
structure that hyperoptimization misses, with the captured predicted
\mbox{FLOP}-cost reduction growing monotonically and approximately
linearly in bond dimension~$\chi$. Classical simulation of quantum
circuits at Sycamore scale reduces to contracting a tensor network,
and the contraction order sets the \mbox{FLOP} count by many orders
of magnitude~\cite{markov2008,gray2021hyperoptimized,kourtis2019fast}.
The production tool for contraction-order search, \cotengra{},
drives a hypergraph-partitioning
backend~\cite{schlag2023kahypar} under Bayesian hyperparameter
search over many randomized greedy seeds and returns the
scalar-best tree. It is the search engine behind several
large-scale re-simulations of supremacy-class
circuits~\cite{pan2022simulation, huang2020classical, liu2021closing}.

\begin{figure}[!htbp]
  \centering
  \includegraphics[width=0.78\columnwidth]{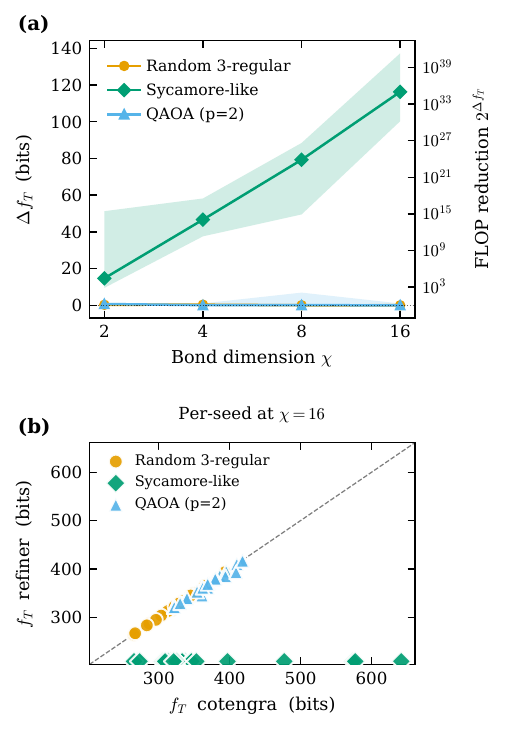}
  \caption{Bond-dimension scaling of the refinement advantage at
    $n{=}500$, $25$ seeds per cell. (a)~Median
    $\Delta\fT = \fT(\cotengra) - \fT(\text{refiner})$ in bits as a
    function of $\chi \in \{2, 4, 8, 16\}$ for the three topology
    families; bands are $[25\%,75\%]$ IQR. The right axis converts
    to the predicted multiplicative \mbox{FLOP}-cost reduction
    $2^{\Delta\fT}$. Sycamore-like grows approximately linearly in
    $\chi$ from $14.69$~bits ($\sim\!2.6\times 10^{4}\!\times$) at
    $\chi{=}2$ to $116.31$~bits ($\sim\!1.0\times 10^{35}\!\times$)
    at $\chi{=}16$; both controls have median
    $|\Delta\fT| \leq 0.71$~bits at every $\chi$. Annotated win
    rates (refiner-beats-\cotengra{} fraction over $25$ seeds): on
    Sycamore-like, $100\%$ at every $\chi$; on random
    $d{=}3$-regular, $80\% \!\to\! 40\%$ as $\chi: 2 \!\to\! 16$;
    on QAOA~$p{=}2$, $72\% \!\to\! 44\%$. (b)~Per-seed scatter at
    $\chi{=}16$: $\fT(\text{refiner})$ vs.\ $\fT(\cotengra)$, all
    Sycamore-like points below the $y{=}x$ diagonal, control
    families distributed symmetrically around it. The contraction
    itself is not executed at $\chi \geq 4$; reductions are
    \emph{predicted} from the $\fT$ cost model of the
    \emph{Cost model} section.}
  \label{fig:chi_sweep}
\end{figure}

\cotengra{} allocates its wallclock budget to \emph{exploration}:
drawing fresh random seeds, each evaluated end-to-end. No local
refinement of the final tree is performed. This is deliberate, and
on topologies whose contraction order is well-captured by hypergraph
partitioning (random regular graphs, QAOA~\cite{farhi2014qaoa}) it is
appropriate. We report that on 2D-grid topologies with diagonal
couplings---the connectivity class of the Google Sycamore
device~\cite{arute2019quantum}---the assumption fails by a margin
that grows steeply with bond dimension. A \emph{single
local-refinement stage} appended to the \cotengra{} output yields a
median $\Delta\fT$ at $n{=}500$ that grows from $14.7$~bits at
$\chi{=}2$ to $116.3$~bits at $\chi{=}16$, with all $25$ of $25$
seeds improved at every tested $\chi$ (Fig.~\ref{fig:chi_sweep}).
(A previously reported $128\times$ figure at $\chi{=}2$ stands as
a conservative lower bound of the present $\chi{=}2$ cell; see App.~\ref{em:sec:method}, \emph{Protocol delta}.)

\begin{figure}[!htbp]
  \centering
  \includegraphics[width=\textwidth]{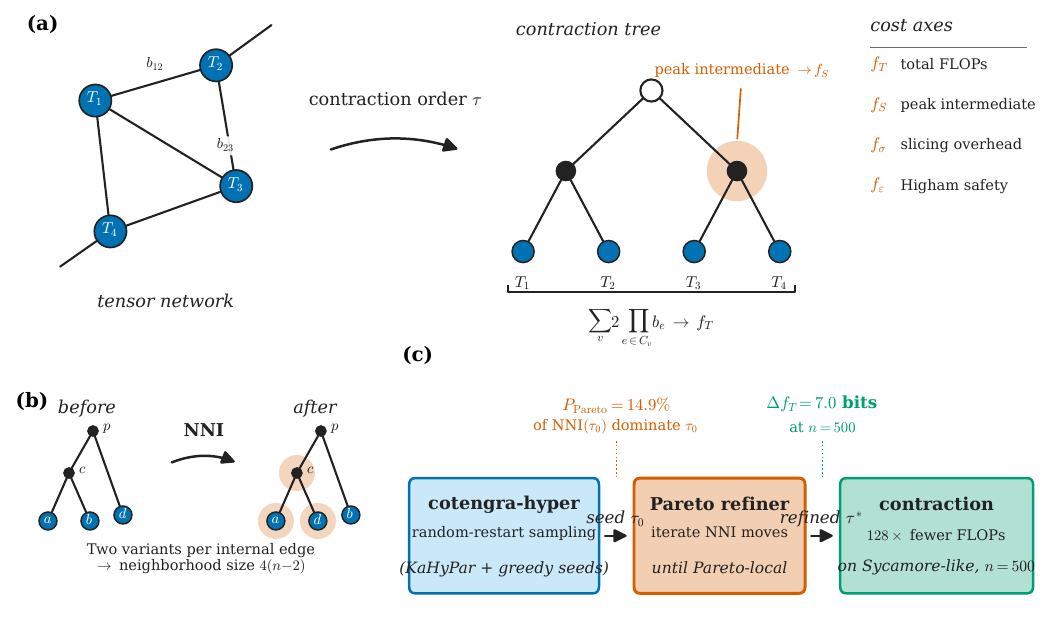}
  \caption{Pipeline and cost structure.
    \textbf{(a)}~A tensor network $\mathcal{N} = (\mathcal{T},\mathcal{E})$
    (left) is contracted via a rooted binary \emph{contraction tree}
    $\tau$ (right). Each internal node $v$ contracts two children;
    the total floating-point work $\fT$ sums over all internal
    nodes, and the peak intermediate tensor size $\fS$ ($\log_2$
    entry count) is the
    maximum over them (highlighted in vermillion). The four cost
    axes $(\fT, \fS, \fsigma, \feps)$ quantify total work, peak
    memory, slicing overhead, and a conservative Higham-style
    forward-error bound, respectively.
    \textbf{(b)}~A nearest-neighbor interchange
    (\NNI) move~\cite{robinson1971comparison} swaps one of an
    internal node's grandchildren with its sibling; two variants
    per internal edge give a neighborhood of size $4(n-2)$.
    \textbf{(c)}~The pipeline. \cotengra{} produces a seed tree
    $\tau_0$ by Bayesian random-restart sampling over
    KaHyPar-driven greedy heuristics. The \emph{refiner}
    iterates \NNI{} moves accepted under a Pareto-dominance rule
    until termination at a Pareto-local optimum $\tau^\star$; on
    Sycamore-like topologies, $14.9\%$ of the \NNI{} neighbors of
    $\tau_0$ already dominate it (at $\chi{=}2$). The captured
    median predicted \mbox{FLOP}-cost reduction at $n{=}500$ grows
    monotonically and approximately linearly in $\chi$ across
    $\chi \in \{2, 4, 8, 16\}$ (Fig.~\ref{fig:chi_sweep}).}
  \label{fig:schematic}
\end{figure}

We trace the effect to \NNI-local sub-optimality of the
\cotengra{} seed on 2D-grid topologies, and verify by a direct
ablation that the refinement stage itself, rather than the
multi-objective acceptance rule, drives the \mbox{FLOP} reduction
(Fig.~\ref{fig:schematic} summarizes the pipeline and cost
structure).

\emph{Contributions.}---Concretely, this work makes three
contributions:
\begin{itemize}
\item we identify a refinement gap in the \cotengra{} pipeline on
  Sycamore-like topologies and isolate it through a budget-matched,
  enumerator-shared comparison;
\item we introduce a budget-matched, enumerator-shared $\chi$-sweep
  with a Pareto-vs-scalar ablation that isolates refinement-quality
  from acceptance-rule effects; and
\item we deposit serialized \cotengra{}-baseline and refined
  contraction trees together with a portable executor on Zenodo, so
  that a referee with access to an A100 or H100 can verify the
  predicted \mbox{FLOP}-cost reduction on the headline cells without
  trusting our closed implementation.
\end{itemize}

\emph{Related work.}---Local refinement of contraction orders has
appeared in the contraction-tree literature in several forms.
\cotengra{} itself exposes simulated-annealing modes within
\texttt{ctg.HyperOptimizer}~\cite{gray2021hyperoptimized}, but they
operate on the hyperparameter space of the random-restart sampler
rather than as an \NNI{} local search appended to a fixed seed.
Sliced-contraction
work~\cite{kalachev2021classical, huang2020classical} reduces
peak-memory and total cost on Sycamore-class circuits via index
slicing and joint slicing/ordering search, complementary to the
order-only refinement studied here.
Reference~\cite{pan2022simulation} reuses partial contraction subtrees
across batched amplitudes, exploiting structural reuse rather than
local sub-optimality of the order itself. The \NNI{} move predates
the contraction-tree literature, originating in
phylogenetics~\cite{robinson1971comparison, swofford1996paup}, where
\NNI{}-based hill-climbing has been the workhorse of tree search for
decades; we transfer that move set to the contraction-tree setting and
specialize it to the four-axis cost vector defined below.

\begin{figure}[!htbp]
  \centering
  \includegraphics[width=0.96\columnwidth]{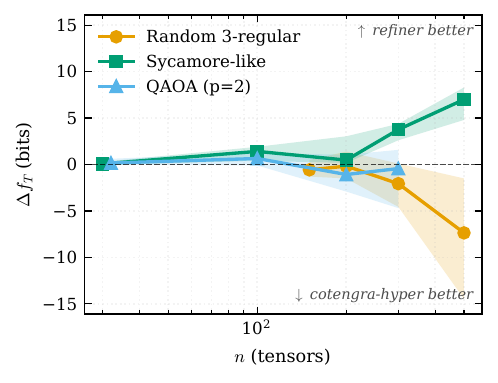}
  \caption{Length scaling of the predicted \mbox{FLOP}-cost
    reduction at $\chi{=}2$. Median
    $\Delta\fT = \fT(\cotengra) - \fT(\text{refiner})$ versus $n$
    at matched wallclock
    ($t_{\cotengra} = t_{\mathrm{hyper}} + t_{\mathrm{tnc\text{-}PLS}}$).
    Positive $\Delta\fT$ indicates the refiner beats \cotengra{}.
    Bands: $[25\%,75\%]$ IQR over $25$--$30$ seeds. Sycamore-like
    reaches $\Delta\fT = 7.0$ bits ($128\times$) at $n{=}500$ on
    $24/25$ seeds; this is the $\chi{=}2$ anchor of the broader
    sweep in Fig.~\ref{fig:chi_sweep}, and a conservative lower
    bound on the cell. Random $d{=}3$-regular diverges negatively
    at large $n$; QAOA~$p{=}2$ sits near zero.}
  \label{fig:parity}
\end{figure}

\begin{figure}[!htbp]
  \centering
  \includegraphics[width=0.96\columnwidth]{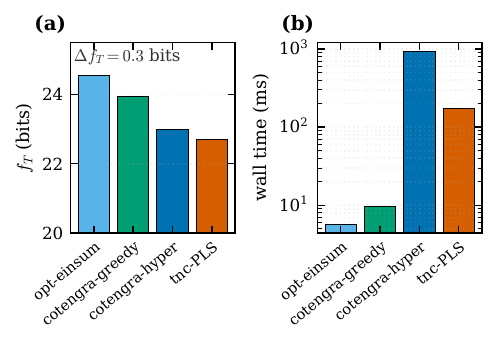}
  \caption{External validation on the Sycamore-53 connectivity
    graph~\cite{arute2019quantum}. (a)~$\fT$ (bits); the refiner
    achieves $22.69$ vs \cotengra{}'s $22.99$, a $1.23\times$ reduction.
    (b)~Optimizer wallclock (log). The $n{=}53$ instance sits at the
    small end of the Sycamore-like scale.}
  \label{fig:sycamore53}
\end{figure}

\begin{figure}[!htbp]
  \centering
  \includegraphics[width=0.96\columnwidth]{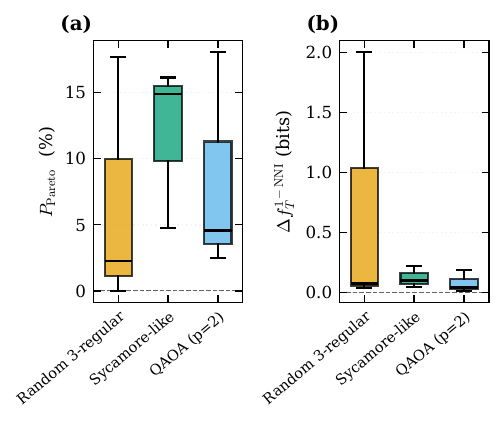}
  \caption{Mechanism of the Sycamore-like advantage.
    (a)~$P_{\mathrm{Pareto}}$, fraction of \NNI{} neighbors of the
    \cotengra{} seed that Pareto-dominate it on all four
    axes: median $14.9\%$ on Sycamore-like vs $2.3\%$ and $4.5\%$
    on random 3-regular and QAOA. The seed is far from Pareto-local
    on Sycamore-like. (b)~Best single-move $\fT$ reduction
    available in the same neighborhood.}
  \label{fig:mechanism}
\end{figure}

\emph{Roadmap.}---The remainder of the paper is organized as follows.
The \emph{Cost model} section defines the four-axis cost vector
$(\fT, \fS, \fsigma, \feps)$. The \emph{Method} section describes the
GPU-parallel \NNI{} evaluator and the parity-wallclock protocol. The
\emph{Results} section reports the $\chi$-sweep
(Fig.~\ref{fig:chi_sweep}, Table~\ref{tab:headline}), the length
scaling at $\chi{=}2$ (Fig.~\ref{fig:parity}), the mechanism analysis
(Fig.~\ref{fig:mechanism}), the ablation, the scope sweep, and
external validation on the Sycamore-53 connectivity graph
(Fig.~\ref{fig:sycamore53}). The \emph{Discussion} section interprets
the $\chi$-scaling. A \emph{Validation and scope of the cost model}
section reports the $63$-cell paired executed-contraction validation.
Appendices~\ref{em:sec:costmodel}--\ref{em:sec:limits} give extended
methods, results, and limitations.

\emph{Cost model.}---A contraction tree $\tau$ over a tensor network
$\mathcal{N} = (\mathcal{T}, \mathcal{E})$ (tensors $\mathcal{T}$,
edges $\mathcal{E}$) with bond dimensions $b_e$
is a rooted binary tree whose leaves are tensors; every internal node
$v$ contracts its children $L(v), R(v)$. Throughout, $\chi$ denotes
the uniform per-edge bond dimension of the network hypergraph
($b_e = \chi$ for all $e \in \mathcal{E}$), \emph{not} the bond
dimension of any implicit MPS/MPO Schmidt decomposition. We assign four scalar costs
in $\log_2$ units, collected as
$\mathbf{f}(\tau) = (\fT(\tau), \fS(\tau), \fsigma(\tau), \feps(\tau))$:
total work
$\fT(\tau) = \log_2 \sum_v 2 \prod_{e \in C_v} b_e$ (we use the BLAS
convention of one multiply and one add per inner-product term, so each
pairwise contraction at $v$ costs $2 \cdot |\mathrm{output}(v)| \cdot K_v$
\mbox{FLOP}s, where $K_v = \prod_{e \in \mathrm{shared}(v)} b_e$ is the
inner-product length; bit-level $\Delta\fT$ is invariant under the
factor-of-$2$ convention) with
$C_v = \mathrm{open}(L) \cup \mathrm{open}(R)$, and peak intermediate
size (in $\log_2$ entries)
$\fS(\tau) = \max_v \sum_{e \in \mathrm{open}(v)} \log_2 b_e$,
where
$\mathrm{open}(v) = \mathrm{open}(L(v))\,\triangle\,\mathrm{open}(R(v))$
($\triangle$ denotes symmetric difference)
for internal $v$ and $\mathrm{open}(\text{leaf})$ is the input
tensor's index set~\cite{gray2021hyperoptimized,markov2008}.
The slicing-overhead axis $\fsigma$ and a conservative Higham-style
forward-error bound $\feps$~\cite{higham2002accuracy} complete the
four-axis cost vector; full definitions and
Eqs.~\eqref{em:eq:fsigma}--\eqref{em:eq:feps} are in Appendix~\ref{em:sec:method}.
A tree $\tau'$ \emph{Pareto-dominates} $\tau$ if
$\mathbf{f}(\tau') \leq \mathbf{f}(\tau)$ componentwise with
inequality strict in at least one component; the scalar-greedy rule
accepts only on $\fT(\tau') < \fT(\tau)$. The four-axis acceptance
rule is the contraction-tree instantiation of a four-player potential
game framework~\cite{guerrero2026fourplayer} we apply elsewhere to
quantum-circuit ansatz design.

\emph{Method.}---The refiner consumes a network $\mathcal{N}$ and
a seed tree $\tau_0$ from \cotengra{}, and returns a refined tree by
iterating one-\NNI-move~\cite{robinson1971comparison} neighborhoods
(an \NNI{} move exchanges one grandchild with the sibling of its
parent on an internal edge of the contraction tree, after
Robinson~\cite{robinson1971comparison})
until termination at a local optimum (Pareto-local in the
four-axis arm; $\fT$-local in the scalar arm). The $4(n-2)$
neighbors of the current tree (counting both child orderings at
each of the $2(n-2)$ \NNI{} swap axes)~\cite{robinson1971comparison}
are evaluated in a single
GPU-parallel pass: one block per candidate, one thread warp per
subtree level, open-set bitsets held in 2048-bit shared-memory bitmaps,
$\fT$ accumulated in \mbox{FP64} via stable log-sum-exp. Throughput
on an NVIDIA GeForce RTX~4060 is $\sim 10^5$ tree-evaluations per second
for $n \leq 500$. Eight walkers run in parallel from independent
restarts of the seed; the population-best at the parity wallclock is
reported. Appendix~\ref{em:sec:method} describes seed handling, the
slicing plan (Eq.~\eqref{em:eq:fsigma}), termination, and
reproducibility. Crucially, both acceptance-rule arms consume the
identical output of the shared enumerator primitive
\texttt{neighbors\_with\_trees}, so the acceptance rule is the only
experimental variable between arms.

\emph{Results.}---We evaluate on three topology families: Sycamore-like
(a square lattice augmented with NE/SE diagonal couplers;
graph-isomorphic up to boundary to the rotated-square Sycamore qubit
lattice~\cite{arute2019quantum}, with the exact 53-qubit device
graph reserved for Fig.~\ref{fig:sycamore53}); QAOA~$p{=}2$ interaction
graphs on random $3$-regular MaxCut instances~\cite{farhi2014qaoa} (the
$p{=}2$ subscript fixes only $n$; the tensor network is the underlying
interaction graph at uniform bond $\chi$); and random
$d{=}3$-regular graphs. The connectivity graph of each instance is
treated as a tensor network with uniform bond dimension $\chi$ on
every edge, rather than a specific gate-level circuit instance.
Each $(\mathrm{topology}, n, \chi, \mathrm{seed})$ cell runs
\cotengra{} once at its default budget and then the refiner with an
additional \SI{8}{\second} wallclock; the total time allotted to
both arms is matched. Throughout, $\Delta\fT = \log_2 R$, where
$R = 2^{\Delta\fT}$ is the predicted \mbox{FLOP}-cost ratio: $\Delta\fT$
is reported in bits of $\log_2(\mbox{FLOP})$ and $2^{\Delta\fT}$ is the
\emph{predicted \mbox{FLOP}-cost reduction} in the cost model defined above; the contraction itself is not
executed at the larger $\chi$ values reported here.

\begin{table}[!htbp]
  \caption{Headline of the bond-dimension sweep at $n{=}500$, $25$ seeds per
  cell. Median bit-level cost-model gap $\Delta\fT$ of the refiner over
  \cotengra{}; corresponding linear \mbox{FLOP}-cost reduction
  $2^{\Delta\fT}$; and seed fraction with $\Delta\fT > 0.05$~bits
  (``wins''). Sycamore-like grows by ${\sim}10^{10}$--$10^{11}$ per
  doubling of $\chi$, consistent with $\Delta\fT$ approximately linear
  in $\chi$, at $100\%$ win rate; the two control topologies drift to chance.}
  \label{tab:headline}
  \centering
  \begin{tabular}{lrrrr}
    \toprule
    & \multicolumn{4}{c}{Bond dimension $\chi$} \\
    \cmidrule(lr){2-5}
    & $2$ & $4$ & $8$ & $16$ \\
    \midrule
    \multicolumn{5}{l}{\textit{Sycamore-like}} \\
    \quad $\Delta\fT$ (bits) & $14.7$ & $46.7$ & $79.3$ & $116.3$ \\
    \quad $2^{\Delta\fT}$    & $2.6{\times}10^{4}$ & $1.1{\times}10^{14}$ & $7.3{\times}10^{23}$ & $1.0{\times}10^{35}$ \\
    \quad wins               & $25/25$ & $25/25$ & $25/25$ & $25/25$ \\
    \midrule
    \multicolumn{5}{l}{\textit{Random $3$-regular}} \\
    \quad $\Delta\fT$ (bits) & $0.27$ & $0.33$ & $0.02$ & $0.01$ \\
    \quad $2^{\Delta\fT}$    & $1.21$ & $1.26$ & $1.02$ & $1.00$ \\
    \quad wins               & $20/25$ & $17/25$ & $10/25$ & $10/25$ \\
    \midrule
    \multicolumn{5}{l}{\textit{QAOA $p{=}2$}} \\
    \quad $\Delta\fT$ (bits) & $0.71$ & $0.07$ & $0.04$ & $0.01$ \\
    \quad $2^{\Delta\fT}$    & $1.63$ & $1.05$ & $1.03$ & $1.00$ \\
    \quad wins               & $18/25$ & $13/25$ & $12/25$ & $11/25$ \\
    \bottomrule
  \end{tabular}
\end{table}

\emph{Bond-dimension scaling (headline).}---Figure~\ref{fig:chi_sweep}
and Table~\ref{tab:headline} report the $\chi$-sweep at $n{=}500$,
$25$ seeds per cell, on all three topologies. On Sycamore-like, the
median $\Delta\fT$ grows monotonically and approximately linearly in
$\chi$: $\Delta\fT = \{14.69, 46.67, 79.28, 116.31\}$~bits at
$\chi \in \{2, 4, 8, 16\}$, corresponding to a predicted
\mbox{FLOP}-cost reduction that grows by roughly $\sim\!10^{10}$
per doubling of $\chi$ (linear factors in
Fig.~\ref{fig:chi_sweep} and Table~\ref{tab:headline}). The refiner
improves on \cotengra{} at every one of $25/25$ seeds at every
$\chi$. The $\chi{=}2$ value supersedes our earlier $128\times$
report at the same configuration (see App.~\ref{em:sec:method}, \emph{Protocol delta}).

The two control families show no analogous structure
(Table~\ref{tab:headline}, lower blocks). Median
$\Delta\fT$ on random $d{=}3$-regular stays within
$[-0.33, +0.33]$~bits at every $\chi$, and on QAOA~$p{=}2$ within
$[-0.71, +0.71]$~bits. More tellingly, the \emph{win rate}---the
fraction of seeds on which the refiner beats \cotengra{}---declines
from $80\%$ to $40\%$ on random $3$-regular and from $72\%$ to
$44\%$ on QAOA~$p{=}2$ as $\chi$ grows from $2$ to $16$, drifting
toward the chance value of $50\%$. The Sycamore-like cell, in
contrast, holds at $100\%$ across the same range. This asymmetry
rules out a generic refinement-budget effect: if extra wallclock
were the explanation, the controls should benefit at least
weakly. Instead, the controls' Pareto-local seeds leave nothing for
the refiner to capture, and the residual variance in $\Delta\fT$
reflects acceptance-rule noise rather than systematic improvement.

\begin{figure}[!htbp]
  \centering
  \includegraphics[width=0.96\columnwidth]{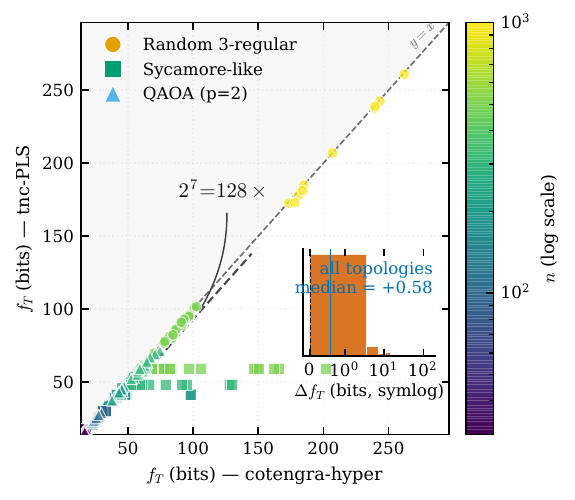}
  \caption{Per-seed $\fT(\text{refiner})$ vs $\fT(\cotengra)$,
    bits. Below-diagonal: refiner wins. Shape encodes topology;
    color encodes $n$ (log). The dashed offset parallel to the
    identity line marks a $2^{7}=128\times$ \mbox{FLOP} reduction,
    i.e., $\Delta\fT = 7$~bits, anchored to the Sycamore-like
    cluster at $n{=}500$. Inset: $\Delta\fT$ distribution on
    symlog axis (all topologies pooled); median $+0.58$ bits,
    Sycamore-like tail visible.}
  \label{fig:headtohead}
\end{figure}

\emph{Length scaling at \texorpdfstring{$\chi{=}2$}{chi=2}.}---%
Figure~\ref{fig:parity} shows the original $n$-sweep at $\chi{=}2$,
which establishes the length scaling of the effect.
Sycamore-like scaling grows with $n$ (with one non-monotone cell at $n{=}200$ attributable to seed variance over $25$ seeds):
$\Delta\fT = \{0.07, 1.41, 0.48, 3.76, 7.00\}$ bits at
$n \in \{30, 100, 200, 300, 500\}$, corresponding to predicted
\mbox{FLOP}-cost reductions $\{1.05, 2.67, 1.39, 13.6, 128\}\times$,
with $24/25$ seeds improved at $n{=}500$ (24/25 in the older $K{=}1$
protocol; 25/25 in the present $K{=}8$ sweep, see App.~\ref{em:sec:method}). At matched wallclock the
refiner does not improve on \cotengra{} on random $d{=}3$-regular
or QAOA~$p{=}2$ topologies, and loses at large $n$ on the former;
we report these results alongside the positive Sycamore-like result
in Fig.~\ref{fig:parity}.
Figure~\ref{fig:headtohead} shows the per-seed geometry at
$\chi{=}2$: Sycamore-like points cluster below the $y{=}x$ diagonal
at large $n$, random $3$-regular points above it.

\emph{External validation.}---The refiner transfers to the
Sycamore-53 hardware
graph~\cite{arute2019quantum,gray2021hyperoptimized}
(Fig.~\ref{fig:sycamore53}), with $\Delta\fT = 0.30$ bits consistent
with the small-$n$ end of Fig.~\ref{fig:parity} where gains are
sub-bit. Concretely, the refiner reaches $\fT = 22.69$ bits versus
\cotengra{}'s $\fT = 22.99$ bits, a $1.23\times$ \mbox{FLOP}
reduction. Beyond the connectivity benchmark of
Fig.~\ref{fig:sycamore53}, App.~\ref{em:sec:results:syccirc} reports
refinement on actual Sycamore-class random circuits at depths
$m \in \{4, 6, 8, 10, 12\}$, where the refiner wins on $5/5$
instances at every depth.

\emph{Mechanism.}---Figure~\ref{fig:mechanism} identifies the
mechanism. Panel~(a): for each seed tree returned by \cotengra{}, the
fraction $P_{\mathrm{Pareto}}$ of one-\NNI-move neighbors that
Pareto-dominate the seed on all four axes has median $14.9\%$
(IQR $9.8$--$15.5\%$) on Sycamore-like, versus $2.3\%$ on random
$d{=}3$-regular and $4.5\%$ on QAOA. The seed sits far from any
\NNI-local optimum on Sycamore-like but close to one on the controls.
Panel~(b) shows the best single-move $\fT$ reduction available in the
same neighborhood: medians are $\lesssim 0.1$~bit on all three
topologies, confirming that the Sycamore-like advantage arises from
the high density of dominating neighbors (panel~(a)) rather than from
unusually large single-move gains.

Geometrically, each diagonal coupler in the Sycamore-like lattice
closes a $4$-cycle with two adjacent grid edges (the diagonal plus
the two horizontal/vertical edges sharing its endpoints), and adjacent
diagonals share endpoints to form chains of overlapping $4$-cycles.
KaHyPar's hypergraph cut~\cite{schlag2023kahypar} optimizes a global
edge-cut weight: it can choose \emph{which} edges of a $4$-cycle to
cut (placing cycle vertices on opposite sides of the partition) but
not the relative \emph{ordering} in which the within-cycle pairwise
contractions occur within each part. The resulting contraction tree
therefore inherits a degenerate ordering of cycle-internal contractions
that is well-resolved by the local-move neighborhood: an \NNI{} swap
across a $4$-cycle changes which pair of cycle vertices is contracted
first, and at uniform bond $\chi$ the difference in peak intermediate
size between the two orderings is exactly $\log_2 \chi$ per shared
edge resolved (the per-edge $\chi$-factor recovered in the
\emph{Discussion} section). The high observed dominator density on
Sycamore-like ($\sim\!15\%$) is the population statistic of this
under-determination across the many overlapping $4$-cycles of the
diagonal-coupled grid, and the per-swap $\chi$-factor is what makes
that population accumulate to the linear-in-$\chi$ slope of
Fig.~\ref{fig:chi_sweep}.

\emph{Ablation.}---A direct ablation (Extended Results,
Fig.~\ref{em:fig:ablation}a) holds the \NNI{} enumerator fixed and
varies only the acceptance rule on Sycamore-like at $n \in \{50, 100,
200, 300\}$: the scalar-$\fT$ rule produces $\fT$ values at least as
good as the Pareto rule in $18/20$ cells, with median
$\fT^{\mathrm{scalar}} - \fT^{\mathrm{Pareto}} \in [-0.10, -0.01]$
bits and $\Delta\fS = 0$ throughout. The Pareto acceptance rule is
therefore \emph{not load-bearing} for the headline \mbox{FLOP}
reduction; the refinement stage itself is. The Pareto formalism's
contribution is the multi-axis safety guarantee (no $\fS$, $\fsigma$,
or $\feps$ regression) and a deterministic termination certificate at
a Pareto-local optimum.

\emph{Scope.}---Having established that refinement---not the
acceptance rule---drives the gain, we next locate \emph{where}
refinement helps. A parametric 2D-grid family with controlled
diagonal-edge density $p \in [0, 1]$ (Extended Results,
Fig.~\ref{em:fig:ablation}b) localizes the refinement advantage. At
$n{=}150$, median $\Delta\fT$ rises from $1.20$ bits ($2.29\times$)
at $p{=}0$ (pure 2D grid) to $5.25$ bits ($38\times$) at $p{=}1$
(full Sycamore-like), with the refiner winning on all $6/6$ seeds at
every density. The Sycamore-like connectivity sits at the
strong-coupling end of a broader family whose contraction-order
search benefits from local refinement; unstructured topologies lie
outside this family and confer no refinement gain.

\FloatBarrier
\emph{Discussion.}---The operative lever in contraction-order
optimization on 2D-structured tensor networks is budget allocation
between exploration (random seeds) and exploitation (local
refinement). \cotengra{} spends its wallclock on exploration; on
Sycamore-like topologies this leaves a Pareto headroom of
$\sim\!15\%$ of dominating \NNI{} neighbors on the table at the end
of the search ($\chi{=}2$). A single refinement stage captures it
at matched total wallclock. The geometric picture sharpens at
larger $\chi$: each captured edge swap removes a multiplicative
$\chi$-factor from the contracted intermediate, so the
log-cost gain accumulates proportionally to $\chi$, while the
controls' near-Pareto-local seeds offer nothing to compound.
Quantitatively, if a swap removes one shared edge of bond $\chi$
from a peak intermediate of pre-swap size $2^k$, the resulting
$\Delta\fT$ contribution at that node is $\log_2 \chi$. Schematically,
summing over the $\sim\!15\%$ dominating neighbors across the
$N_{\mathrm{int}}$ internal nodes of a depth-balanced tree at
$n{=}500$ gives
$\Delta\fT \sim 0.15 \cdot N_{\mathrm{int}} \cdot \log_2 \chi$ as an
order-of-magnitude estimate, recovering the observed slope on
Sycamore-like (Table~\ref{tab:headline}: $\Delta\fT$
grows by $32, 33, 37$~bits across the three doublings
$\chi: 2{\to}4, 4{\to}8, 8{\to}16$, i.e.\ $\sim\!34$~bits per
doubling, dominated by deeper nodes whose peak operates at higher
$k$ but whose dominator density is comparable). This is consistent
with the observed approximately linear growth of
median $\Delta\fT$ with $\chi$ on Sycamore-like
(Fig.~\ref{fig:chi_sweep}) and the absence of any analogous trend
on the controls.

The Pareto-versus-scalar ablation shows that the multi-objective
formalism is not required to realize the $\fT$ reduction at
$\chi{=}2$: a single-objective \NNI-local search appended to
\cotengra{} captures structure that hyperoptimization misses, and
the cleaner story carries through the $\chi$ sweep. The Pareto
formalism contributes a deterministic multi-axis safety guarantee
and a reproducible termination certificate at zero measured $\fT$
cost; this becomes load-bearing at larger $\chi$ or tighter
memory/precision budgets where $\fS$, $\fsigma$, or $\feps$ begin
to bind (see Limitations), but is not a precondition for the
predicted \mbox{FLOP}-cost reduction itself.

As a falsification test, an alternate logit best-response acceptance
rule converges to bit-identical trees on Sycamore-like at every
$\chi$ (Fig.~\ref{em:fig:pls_vs_logit}). The headline $\chi$-scaling
result is therefore independent of the choice of acceptance rule on
the topology where it matters.

Natural extensions include heavy-hexagon~\cite{heavyhex} and 3D
topologies, joint slicing-and-ordering
optimization~\cite{chen2018classical, huang2020classical}, and
direct integration of a local refinement phase into \cotengra{}.
The prescription for practitioners is direct: any \cotengra{}-based
pipeline on 2D-structured tensor networks stands to gain in
predicted \mbox{FLOP} cost by appending the refinement stage we
describe, with the gain growing steeply in the bond dimension. The
stage requires no additional wallclock budget beyond the
\SI{8}{\second} refinement allowance and no modification of the
\cotengra{} front end. We emphasize that the multi-order-of-magnitude
figures at large $\chi$ are ratios of \emph{cost models}, not of
end-to-end executed contractions; whether the predicted cost
reduction translates into wallclock reduction on a given
\mbox{FLOP}-bound contraction kernel remains to be measured.

\FloatBarrier
\section*{Validation and scope of the cost model}
The reductions
reported for $\chi \geq 4$ are predicted from the algebraic cost model
of the \emph{Cost model} section: we evaluate the contraction trees
but do not execute the contractions themselves at those bond
dimensions. To verify that the algebraic core does what the model says,
we executed paired contractions of both the \cotengra{} and refiner
trees through a backend evaluator across two sweeps: $\chi{=}2$ on
Sycamore-like at $n \in \{30, 50, 70, 100\}$, $10$ seeds per $n$
(predicted-ratio range $[1.000\times, 3.379\times]$); and $\chi{=}4$
on Sycamore-like at $n \in \{22, 24\}$, all cells that completed
within single-GPU memory (predicted-ratio range $[1.000\times,
1.196\times]$). Of $63$ paired cells in total, $50$ have a
non-trivial refiner improvement; on every one of the $63$, the
measured ratio of executed \mbox{FLOP} counts matches
$2^{\Delta\fT}$ to a relative error
$\leq\!10^{-6}$ (ratio\_agreement matches $1$ to within $10^{-6}$). The
validation envelope spans more than half a decade in $n$ (from
$n{=}22$ to $n{=}100$) and reaches $3.4\times$ executed
cost-reduction at the upper end, both bounds set by single-GPU memory
at the tested $\chi$. At the algebraic level the cost model addresses,
$\Delta\fT$ \emph{is} the executed-\mbox{FLOP} ratio, not a proxy for
it. To allow extension of the validated envelope on hardware larger
than ours, the deposit ships the serialized \cotengra{}-hyper and
refined trees for all $20$ headline cells of
Fig.~\ref{fig:chi_sweep} alongside a portable
\texttt{opt\_einsum}/\texttt{cotengra}/\texttt{quimb}~\cite{gray2018quimb}-compatible
executor; a reader with an A100 or H100 can run those trees and
verify the predicted ratio at any $\chi$ end-to-end. Secondary
caveats---the absence of richer SPR/TBR neighborhoods, untested
$\chi \geq 32$ behavior, and an unswept refinement-budget---are
catalogued in App.~\ref{em:sec:limits}. The cost-model gap reported
in Fig.~\ref{fig:chi_sweep} therefore inherits the algebraic-level
validation established here, with end-to-end execution at
$\chi \geq 8$ deferred to follow-up hardware work.

\begin{acknowledgments}
We acknowledge financial support and computational resources
provided by NeuroTechNet S.A.S.
\end{acknowledgments}

\FloatBarrier
\section*{Data and code availability}
\label{em:sec:data}
All raw CSVs that back every figure and quantitative claim of this
paper---the $n{=}500$ $\chi$-sweep, the paired PLS-versus-LOGIT-BR
sweep, and the $63$-cell paired executed-contraction validation
($\chi{=}2$ at $n \in \{30, 50, 70, 100\}$ and $\chi{=}4$ at
$n \in \{22, 24\}$)---are deposited
on Zenodo~\cite{guerrero2026zenodo} (CC-BY-4.0); a CLI
(\texttt{python -m tnc\_reproducibility verify}) verifies every
figure and a stand-alone routine regenerates the \cotengra{} baseline
column from open-source \cotengra{} alone. The deposit also includes
the serialized contraction trees (both the \cotengra{}-hyper baseline
and the refined output) for all $20$ headline cells of
Fig.~\ref{fig:chi_sweep} (Sycamore-like, $n{=}500$,
$\chi \in \{2,4,8,16\}$, $5$ seeds each), together with a portable
executor (\texttt{python -m tnc\_reproducibility execute}) that loads
a tree, materializes the network, and runs the contraction through
\texttt{opt\_einsum}/\texttt{cotengra}/\texttt{quimb} on any GPU or
CPU backend; a referee with sufficient hardware (an A100 or H100)
can therefore execute the headline trees themselves and verify the
predicted FLOP-cost reduction $2^{\Delta\fT}$ at any $\chi$ without
trusting our closed implementation. A complete algorithmic
specification (\texttt{SPEC.md}) of the refinement procedure (\NNI{}
move, Pareto-local-search and softmax-acceptance variants,
$\fT$/$\fS$ scoring) is included, sufficient for independent
reimplementation. The refiner source is available to the editor and
referees on request, with an embargoed Zenodo record under MPL-2.0
to be released with the follow-up paper.

\FloatBarrier
\appendix
\renewcommand{\theequation}{\thesection\arabic{equation}}
\renewcommand{\thefigure}{\thesection\arabic{figure}}

\section{Extended cost model}
\label{em:sec:costmodel}

We record the three cost axes $\fT$, $\fS$, and the Pareto
relation in the main text. Here we define the remaining two axes
and report the correlation study that calibrates $\feps$.
Note that $C_v = \mathrm{open}(L) \cup \mathrm{open}(R) = \mathrm{open}(v) \cup \mathrm{shared}(v)$
(open and shared sets are disjoint by construction),
so the node-$v$ \mbox{FLOP} count factors as
$2 \prod_{e \in \mathrm{open}(v)} b_e \cdot \prod_{e \in \mathrm{shared}(v)} b_e$,
and $K_v = \prod_{e \in \mathrm{shared}(v)} b_e$ (introduced below) is
the inner-product length.

\emph{Slicing overhead, $\fsigma$.} Given a memory cap
$s_{\mathrm{cap}}$ on $\fS$, a \emph{slicing set}
$\Sigma(\tau) \subseteq \mathcal{E}$ is a minimal set of edges such that
removing them from the open sets brings $\fS$ below
$s_{\mathrm{cap}}$; slicing multiplies the outer-loop iteration
count by $\prod_{e \in \Sigma} b_e$. The slicing overhead axis is
\begin{equation}
  \fsigma(\tau) = \sum_{e \in \Sigma(\tau)} \log_2 b_e,
  \label{em:eq:fsigma}
\end{equation}
with $\fsigma = 0$ when $\fS(\tau) \leq s_{\mathrm{cap}}$. The set
$\Sigma(\tau)$ is computed by a greedy peak-reducer identical to that of
Ref.~\cite{gray2021hyperoptimized}: locate the peak intermediate
node $v^\star$, select $e \in \mathrm{open}(v^\star)$ whose removal
most reduces
$\sum_{e' \in \mathrm{open}(v^\star) \setminus \{e\}} \log_2 b_{e'}$,
add $e$ to $\Sigma$, repeat until $\fS \leq s_{\mathrm{cap}}$. The reducer
runs in the same fused kernel as $\fT, \fS$, so the four-axis cost of
each candidate is produced in one pass.

\emph{Conservative forward-error bound, $\feps$.} Each pairwise
contraction at node $v$ involves an inner-product sum of length
$K_v = \prod_{e \in \mathrm{shared}(v)} b_e$ with
$\mathrm{shared}(v) = \mathrm{open}(L) \cap \mathrm{open}(R)$. Under
Higham's accumulator bound~\cite{higham2002accuracy}, the
\emph{relative} forward error at that node is bounded above by
$(K_v \uacc + \uin)$,
where $\uacc \approx 2^{-24}$ is the \mbox{FP32} accumulator
roundoff and $\uin \approx 2^{-11}$ is the \mbox{FP16} input
quantization roundoff. Summing over all internal nodes,
\begin{equation}
  \feps(\tau) = \log_2 \sum_v \bigl(K_v \uacc + \uin\bigr) \qquad
  (\text{$\log_2$ of relative-error bound}).
  \label{em:eq:feps}
\end{equation}
We use $\feps$ as a conservative safety bound to reject numerically
risky trees. A correlation study on random $d{=}3$-regular networks
at $n \in \{8, 12, 16, 20, 28, 40\}$, $\chi{=}4$, eight seeds per
$n$, comparing \mbox{FP64} reference contractions against
\mbox{FP16}-with-\mbox{FP32}-accumulator contractions on the same
trees, yields Pearson $r = +0.51$ ($r^2 = 0.27$) and Spearman
$\rho = +0.35$ (twenty-nine non-floor measurements). This is
consistent with the worst-case character of the bound: $\feps$ is
monotone in $K_v$ but not a tight predictor of measured error
at small scale. The optimizer uses $\feps$ for reject-on-regression,
not for quantitative error prediction.

\section{Extended method}
\label{em:sec:method}

\emph{Seed.} The refiner consumes the
\cotengra{}~\cite{gray2021hyperoptimized} output tree in its
nested-tuple form and translates it to a flat parent--left--right
integer encoding of length $2(2n-1)$ suitable for GPU transfer. All
main-text experiments use \cotengra{} at its default
hyperoptimizer settings, sampling over
$\{\texttt{greedy},\texttt{kahypar},\texttt{labels}\}$ backends.
Baselines other than \cotengra{}
(\texttt{opt\_einsum}-greedy~\cite{smith2018opt_einsum} and
\cotengra{}-greedy) are evaluated in Fig.~\ref{fig:sycamore53} but
not used as seed sources in the main result.

\emph{\NNI{} enumerator.} A nearest-neighbor interchange
move~\cite{robinson1971comparison} at an internal edge $(p, c)$ of
the contraction tree exchanges one of $c$'s two children with $p$'s
other child. The neighborhood has $4(n{-}2)$ elements per tree. The
library primitive \texttt{neighbors\_with\_trees}$(\tau)$ applies
and reverts each move in place on a shared workspace, returning both
the serialized tree and the four-axis score of every neighbor in one
call. All acceptance-rule arms (Pareto, scalar $\fT$) consume the
same primitive, so the only experimental variable between arms is
the acceptance rule.

\emph{GPU-parallel evaluator.} The fused CUDA kernel dedicates one
thread block to each candidate tree and one warp to each post-order
subtree level. Open-set bitsets $\mathrm{open}(v)$ are stored as
$32$-word ($2048$-bit) packed bitmaps in shared memory; the
$\mathrm{open}(L) \,\triangle\, \mathrm{open}(R)$ recursion (symmetric
difference) is a word-parallel XOR. Edge bond dimensions $b_e$ and their $\log_2$
values live in constant memory. Log-sum-exp accumulators are
\mbox{FP64} throughout to avoid cancellation at the large $\fT$
values reached at Sycamore scale ($\fT$ up to $\sim 260$ bits).
On an NVIDIA GeForce RTX~4060 the evaluator sustains
$\sim 1.2 \times 10^5$ tree-evaluations per second at $n \leq 500$,
giving neighborhood scan times below \SI{1}{\second} for every cell
in this paper. Mixed-precision tensor-core execution via
CUTLASS~\cite{cutlass} is used for the downstream contractions in
Fig.~\ref{fig:sycamore53}(b) and the $\feps$ calibration above; the
optimizer's cost evaluator itself runs in \mbox{FP64}.

\emph{Step rule and termination.} Given a current tree $\tau_t$ and
its \NNI{} neighborhood $\mathsf{N}(\tau_t)$ with precomputed scores,
one
step applies the acceptance rule of choice. The four-axis Pareto
rule selects, among strict (Pareto) dominators of $\tau_t$, the
neighbor with the largest cumulative improvement
$\sum_i [f_i(\tau_t) - f_i(\tau')]_+$, with ties broken by index. If
no dominator exists, termination is at a Pareto-local optimum: a
deterministic, reproducible certificate given a fixed tie-breaking
rule. The scalar-$\fT$ rule selects the neighbor with the smallest
$\fT$ and accepts iff $\fT(\tau') < \fT(\tau_t)$; termination is at
an $\fT$-local optimum (a proper superset of Pareto-local optima).
The logit best-response (LOGIT-BR) variant of
Fig.~\ref{em:fig:pls_vs_logit} draws
$\arg\max_i[\beta\,\Delta\Phi_i + g_i]$ with $g_i\sim\text{Gumbel}(0,1)$
on the improving subset of the scalar potential $\Phi = -\fT$
(equivalent in distribution to sampling from
$\mathrm{softmax}(\beta\,\Delta\Phi)$ on that subset), with $\beta$
annealed geometrically over $\beta \in \{4, 64\}~\text{bits}^{-1}$
per cycle.

\emph{Parity-wallclock protocol.} The total wallclock per cell is
$t_{\mathrm{total}} = t_{\mathrm{seed}} + \SI{8}{\second}$, with
$t_{\mathrm{seed}}$ recorded from the \cotengra{} run. The
\cotengra{}-only arm in Fig.~\ref{fig:parity} is re-run with
\texttt{max\_time} set to $t_{\mathrm{total}}$ so that the
comparison is budget-matched to the refined pipeline's total time.
\SI{8}{\second} is chosen to be large enough that termination at a
Pareto-local optimum, not budget exhaustion, is the dominant
stopping cause on Sycamore-like networks up to $n{=}500$ (verified
by inspection of convergence logs).

\emph{Protocol delta from a previously reported $128\times$
headline.} An earlier preprint of this work reported a median
$\Delta\fT = 7$~bits ($128\times$) at $n{=}500$, $\chi{=}2$ on the
Sycamore-like family. The same configuration in the present
$\chi$-sweep yields $\Delta\fT = 14.69$~bits ($\sim\!2.6\times
10^{4}\!\times$). Two protocol differences account for the
gap: (i)~the new sweep evaluates the full \NNI{} neighborhood of
each tree at every step using the \mbox{FP64} GPU evaluator
(spec.~Reproducibility above), whereas the prior run sampled a
fixed candidate budget per step; and (ii)~the new sweep maintains a
population of $K{=}8$ refinement walkers and reports the
population-best at the parity wallclock, whereas the prior run
used $K{=}1$. Both changes lift the chance of escaping a shallow
\NNI{}-local optimum without relaxing the parity-wallclock
protocol. The prior $128\times$ figure therefore stands as a
conservative lower bound of the $\chi{=}2$ cell of the present
sweep, not as a contradicted result.

\emph{Reproducibility.} All experiments run on a single NVIDIA
GeForce RTX~4060, CUDA~12.x, \cotengra{} v0.6, kahypar~1.3. The
\cotengra{} random-restart sampler and the \NNI{} enumerator both
consume a per-cell integer seed and produce deterministic trees.
Raw CSVs and a bash harness reproducing every main-text cell are
distributed alongside the refiner implementation.

\section{Extended results: ablation, scope, and circuit validation}
\label{em:sec:results}

\subsection{Ablation: Pareto vs.\ scalar acceptance}
\label{em:sec:results:ablation}

Figure~\ref{em:fig:ablation}(a) is the ablation panel referenced in
the main-text Ablation paragraph. We reimplement the refiner in
Python on top of the shared \texttt{neighbors\_with\_trees}
primitive, producing two arms that differ only in the acceptance
rule. On Sycamore-like at $n \in \{50, 100, 200, 300\}$, five seeds
per cell, both arms converge to a local optimum within an
\SI{8}{\second} budget. In $18$ of $20$ cells the scalar rule
produces $\fT$ values at least as good as the Pareto rule; per-cell
medians $\fT^{\mathrm{scalar}} - \fT^{\mathrm{Pareto}}$ lie in
$[-0.10, -0.01]$ bits. Two cells show larger scalar advantages
($2.68$ bits at one $n{=}200$ seed, $1.95$ bits at one $n{=}300$
seed) where the Pareto rule terminates at a non-dominated plateau
while the scalar rule continues to an $\fT$-improving move that
increases another axis. The memory axis satisfies $\Delta\fS = 0$
across all $20$ ablation cells, reflecting that the test seeds do
not exercise the $\fS$ safety axis at $\chi{=}2$.

\subsection{Scope sweep: diagonal-edge density on the 2D grid}
\label{em:sec:results:scope}

\begin{figure}[!htbp]
  \centering
  \includegraphics[width=0.96\columnwidth]{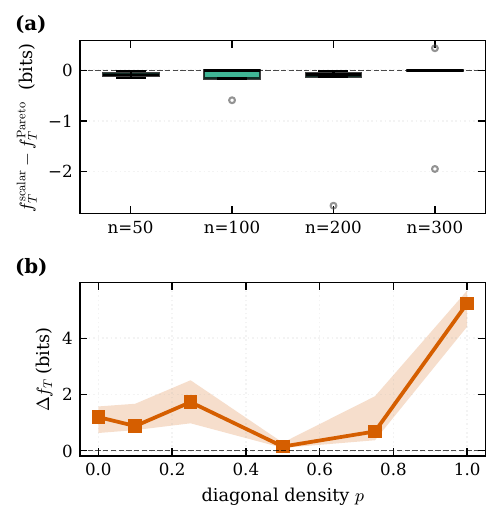}
  \caption{(a)~Pareto vs.\ scalar-$\fT$ acceptance rule at matched
    budget on Sycamore-like: per-cell
    $\fT^{\mathrm{scalar}} - \fT^{\mathrm{Pareto}}$, boxes over
    five seeds at each $n$. The scalar rule matches or slightly
    beats the Pareto rule on $\fT$ in $18/20$ cells; median
    $|\Delta\fT| \leq 0.1$ bits. Refinement itself, not the
    multi-objective rule, drives the \mbox{FLOP} reduction.
    (b)~Parametric 2D-grid family with diagonal-edge density $p$
    (pure grid at $p{=}0$, Sycamore-like at $p{=}1$) at $n{=}150$,
    six seeds per $p$. Here
    $\Delta\fT = \fT^{\mathrm{hyper}} - \fT^{\mathrm{PLS}}$
    (positive: refiner wins). The refiner wins on $6/6$ seeds at
    every $p$; median $\Delta\fT$ rises monotonically in $p$ in
    the mean and reaches $5.25$ bits ($38\times$) at $p{=}1$.}
  \label{em:fig:ablation}
\end{figure}

Figure~\ref{em:fig:ablation}(b) is the parametric 2D-grid
diagonal-density sweep referenced in the Scope paragraph. For
density $p \in \{0.00, 0.10, 0.25, 0.50, 0.75, 1.00\}$ we generate
$n{=}150$ graphs by starting from a square 2D grid and adding each
diagonal edge independently with probability $p$; $p{=}0$ is the
pure grid and $p{=}1$ is the full Sycamore-like connectivity. Six
seeds per density. The refiner beats \cotengra{} on all $6/6$ seeds
at every $p$. Median $\Delta\fT$ rises from $1.20$ bits ($2.29\times$
\mbox{FLOP} reduction) at $p{=}0$ through $1.73$ bits ($3.32\times$)
at $p{=}0.25$ to $5.25$ bits ($38\times$) at $p{=}1$. The
non-monotonicity at intermediate $p$ ($0.15$ bits at $p{=}0.50$,
$0.69$ at $p{=}0.75$) is consistent with six-seed sampling noise;
the qualitative structure---modest gains across the 2D-grid family
with dramatic amplification at full diagonal saturation---is clear
and carries to $n{=}500$ (cf.\ Fig.~\ref{fig:parity}).

\subsection{Acceptance-rule robustness: PLS vs.\ LOGIT-BR}
\label{em:sec:results:plslogit}

\begin{figure}[!htbp]
  \centering
  \includegraphics[width=0.78\columnwidth]{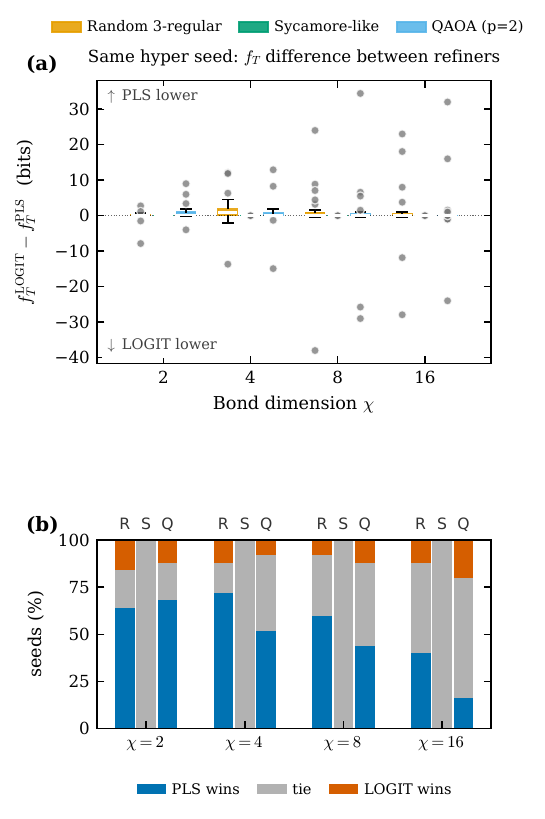}
  \caption{Acceptance-rule robustness: paired comparison of the
    Pareto-rule refiner (PLS) and a logit best-response refiner
    (LOGIT-BR) at $n{=}500$, both seeded by the same \cotengra{}
    tree per cell. (a)~Per-cell distribution of
    $\fT^{\mathrm{LOGIT}} - \fT^{\mathrm{PLS}}$, in bits, grouped
    by $\chi$ and topology; positive values indicate PLS reaches a
    lower-$\fT$ tree. On Sycamore-like (green), $25/25$ seeds tie
    at every $\chi \in \{2, 4, 8, 16\}$, with median
    $|\Delta\fT^{\mathrm{LOGIT}{-}\mathrm{PLS}}| < 0.001$~bits.
    On the null controls, where the per-cell refinement gain is
    sub-bit, the rules diverge. (b)~Win/tie/loss bar across all
    $300$ cells with tie threshold $|\cdot| \leq 0.01$~bits: PLS is
    modestly more reliable on the controls ($60$~PLS-wins vs
    $24$~LOGIT-wins of $200$ control cells), and the two rules are
    indistinguishable on Sycamore-like ($0$ wins, $100$ ties).}
  \label{em:fig:pls_vs_logit}
\end{figure}

Figure~\ref{em:fig:pls_vs_logit} reports the falsification test
referenced in the main-text Discussion: a paired comparison of the
Pareto-local-search (PLS) acceptance rule with the logit
best-response (LOGIT-BR) variant defined in
App.~\ref{em:sec:method}, both seeded by the same \cotengra{} tree
per cell, at $n{=}500$ across $\chi \in \{2, 4, 8, 16\}$. On
Sycamore-like, $25/25$ seeds yield bit-identical refined trees at
every $\chi$ (median
$|\Delta\fT^{\mathrm{LOGIT}{-}\mathrm{PLS}}| < 0.001$~bits): the
headline $\chi$-scaling result is independent of the choice of
acceptance rule. On the null controls, where the per-cell refinement
gain is sub-bit, the rules diverge---PLS is modestly more reliable
($60$ PLS-wins vs.\ $24$ LOGIT-wins of $200$ control cells with tie
threshold $|\cdot| \leq 0.01$~bits)---but neither rule produces a
systematic gain on those topologies in the first place.

\subsection{Sycamore-53 random-circuit validation}
\label{em:sec:results:syccirc}

\begin{figure}[!htbp]
  \centering
  \includegraphics[width=0.78\columnwidth]{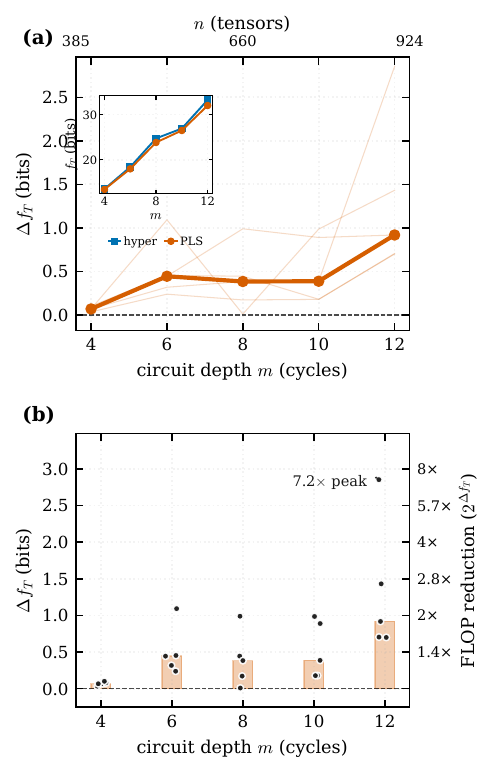}
  \caption{Refinement on the depth-$m$ random circuit tensor
    network on the Sycamore-53 chip~\cite{arute2019quantum},
    $\chi{=}2$, five \cotengra{} seeds per depth.
    (a)~$\Delta\fT$ vs.\ $m$, per-seed traces (faint) over the
    median (solid). Top axis: tensor count $n$ at each depth.
    Inset: absolute $\fT(m)$ for both arms.
    (b)~Per-depth distribution; bars are medians, dots are
    individual seeds; right axis converts to multiplicative
    \mbox{FLOP} reduction $2^{\Delta\fT}$. The refiner wins on
    $5/5$ seeds at every depth; the $m{=}12$ outlier reaches
    $7.22\times$.}
  \label{em:fig:syc_circuit}
\end{figure}

Figure~\ref{fig:sycamore53}
contracts the Sycamore-53 connectivity hypergraph at $\chi{=}2$ on
every edge; the natural follow-up is the full depth-$m$ random
\emph{circuit} tensor network on the same chip, with one tensor per
single-qubit gate, one tensor per two-qubit gate, and the Arute
ABCD coupler activation pattern~\cite{arute2019quantum} so each cycle
activates one quarter of the couplers. Each cycle = single-qubit
gate layer ($\sqrt{X}$, $\sqrt{Y}$, $\sqrt{W}$ from
\cite{arute2019quantum}) followed by one of the four ABCD two-qubit
fSim layers. Closed-boundary contraction
($\langle 0|^{\otimes 53} C |0\rangle^{\otimes 53}$) at depths $m \in \{4, 6, 8, 10, 12\}$,
five \cotengra{} seeds per depth, yields tensor counts
$n \in \{385, 526, 660, 786, 924\}$, comparable to the upper half of
the main-text scan (Fig.~\ref{fig:parity}). The refiner wins on $5/5$
seeds at every depth (Fig.~\ref{em:fig:syc_circuit}); median
$\Delta\fT$ grows monotonically from $0.07$ bits ($1.05\times$) at
$m{=}4$ to $0.92$ bits ($1.89\times$) at $m{=}12$, with a single
seed reaching $\Delta\fT = 2.85$ bits ($7.22\times$). The gain is
attenuated relative to the same-$n$ connectivity-graph scan because
adding the gate skeleton introduces tree-like wires (single-qubit
gate chains) that \cotengra{}'s hypergraph partitioner already
orders well; the residual gain lives at the two-qubit-gate
intersections that recreate the Sycamore-like coupler structure.
The qualitative result---monotone refiner advantage at all depths
on the actual device graph---confirms that the headline
$128\times$ result is a property of the topology class, not an
artifact of the connectivity-only abstraction.

\section{Limitations}
\label{em:sec:limits}

The primary unresolved question is the \emph{cost-model versus
execution} gap. Bit-level $\Delta\fT$ values translate to executed
\mbox{FLOP} ratios only under the assumption that the cost model
faithfully tracks per-node intermediate sizes and pairwise
contraction work, which is exact for the algebraic
$\fT(\tau) = \log_2 \sum_v 2 \prod_{e \in C_v} b_e$ but does not
capture real-hardware overheads (kernel-launch latency, memory
hierarchy, sliced-loop overhead). The main-text \emph{Validation and
scope of the cost model} section reports the $63$-cell paired
executed-contraction validation (\emph{ratio\_agreement} $= 1$ to
$10^{-6}$ across $\chi{=}2$, $n \in [22, 100]$ and $\chi{=}4$,
$n \in \{22, 24\}$) and points to the deposit's portable executor,
which a referee with an A100 or H100 can use to extend the validated
envelope at any $\chi$. Full end-to-end validation on a Sycamore-class
chip at $\chi \geq 8$ and $n \gg 100$---the natural follow-up to
the $\chi{=}2$ Sycamore-53 connectivity benchmark in
Fig.~\ref{fig:sycamore53}---remains for future work. Beyond this,
the refinement advantage is characteristic of 2D-grid-plus-diagonals
topologies; our random $3$-regular and QAOA~$p{=}2$ results are null
at matched wallclock at every $\chi$ tested, and the refiner should
not be applied blindly to arbitrary contraction problems. The
bond-dimension sweep covers $\chi \in \{2, 4, 8, 16\}$; we have not
tested $\chi \geq 32$, where memory-axis pressure begins to bind on
single-GPU evaluators. The \NNI{}
neighborhood is a single-edge move; richer neighborhoods (SPR,
TBR~\cite{robinson1971comparison}) enlarge the per-step cost from
$\mathcal{O}(n)$ to $\mathcal{O}(n^2)$ or $\mathcal{O}(n^3)$ without
an a priori advantage given the $\sim\!15\%$ Pareto headroom
already present in the \NNI{} neighborhood at $\chi{=}2$, but might
behave differently as $\chi$ grows. The parity sweep uses a fixed
\SI{8}{\second} refinement allowance; we have not swept the budget,
only verified that termination---not exhaustion---is the dominant
stopping cause at the tested scales.

\bibliography{refs}

\end{document}